\documentclass[review]{elsarticle}
\usepackage{graphics}
\usepackage{graphicx}
\usepackage{float}
\usepackage{bm}
\usepackage[nomarkers,notablist,nofiglist]{endfloat}
\usepackage{lineno,hyperref}
\modulolinenumbers[5]
\usepackage{siunitx}         
\graphicspath{ {./Figures/} } 
\usepackage{soul}
\usepackage{amsmath}
\usepackage{geometry} 
\usepackage{color} 
\usepackage{outlines}
\journal{arXiv}

\newcommand\blfootnote[1]{%
  \begingroup
  \renewcommand\thefootnote{}\footnote{#1}%
  \addtocounter{footnote}{-1}%
  \endgroup
}

\begin{document}
\begin{frontmatter}
\title{Graph Neural Network Modeling of Grain-scale Anisotropic Elastic Behavior using Simulated and Measured Microscale Data}
\author{Darren C. Pagan\textsuperscript{1*}\blfootnote{*Corresponding Author}}
\author{Calvin R. Pash\textsuperscript{1}}
\author{Austin R. Benson\textsuperscript{2}}
\author{Matthew P. Kasemer\textsuperscript{3}}
\address{\textsuperscript{1}Pennsylvania State University, University Park, PA 16802 USA}
\address{\textsuperscript{2}Cornell University, Ithaca, NY 14853 USA}
\address{\textsuperscript{3}University of Alabama, Tuscaloosa, AL 35487 USA}

\begin{abstract}
Here we assess the applicability of graph neural networks (GNNs) for predicting the grain-scale elastic response of polycrystalline metallic alloys. Using GNN surrogate models, grain-averaged stresses during uniaxial elastic tension in Low Solvus High Refractory (LSHR) Ni Superalloy and Ti 7wt\%Al (Ti-7Al), as example face centered cubic and hexagonal closed packed alloys, are predicted. A transfer learning approach is taken in which GNN surrogate models are trained using crystal elasticity finite element method (CEFEM) simulations and then the trained surrogate models are used to predict the mechanical response of microstructures measured using high-energy X-ray diffraction microscopy (HEDM). The performance of using various microstructural and micromechanical descriptors for input nodal features to the GNNs is explored through comparisons to traditional mean-field theory predictions, reserved full-field CEFEM data, and measured far-field HEDM data. The effects of elastic anisotropy on GNN model performance and outlooks for extension of the framework are discussed.
\end{abstract}

\end{frontmatter}

\section{Introduction}

The local micromechanical response of grains embedded within a polycrystal is dictated not only by an isolated grain's features (e.g., defect state or crystallographic lattice orientation), but also by the features and mechanical response of adjoining grains (i.e., the local grain `neighborhood'). Various linking hypotheses or methods in which grains are embedded within homogeneous matrices---most notably the Eshelby method \cite{eshelby1957determination}---attempt to capture the average features of grain-scale response, but, by construction, do not consider variation of behavior created by the interactions of specific grains within their local neighborhoods. The relatively recent ability to explicitly model grains and grain neighborhoods in three-dimensional (3D) polycrystals using both finite element \cite{marin_a} and spectral \cite{lebensohn2012elasto} methods has allowed these neighborhood effects to be more thoroughly explored. However, the use of these methods comes at a sometimes-significant computational cost. Increasing computational complexity, particularly with increases in microstructural fidelity or the inclusion of various plastic deformation mechanisms in modeling efforts, limits the number of microstructural configurations that can be tested, which consequently limits the ability of these models to be embedded within larger scale simulations (considering current computational capabilities). To address these challenges, low-computational-cost surrogate models which can rapidly evaluate micromechanical response and evaluate large regions of microstructural parameter space are necessary. In this work we propose and demonstrate that the fundamental network (or graph) structure of polycrystals make them candidates for surrogate mechanical modeling through graph neural networks (GNNs)~\cite{hamilton2020graph,wu2020comprehensive,zhou2020graph}. We demonstrate the utility of this surrogate modeling with GNNs trained to predict grain-scale elastic response in two example alloy systems. The GNNs are trained with microscale crystal elasticity finite element method (CEFEM) simulations, and then tested against grain-scale elastic response measured using high-energy X-ray diffraction microscopy (HEDM).

Supervised machine learning has become a primary choice for generating surrogate models for predicting the mechanical properties and performance of engineering alloys \cite{Fuhg2022}. At the microscale, multiple efforts have utilized convolutional neural networks (CNNs) to predict deformation fields at sub-grain length scales in virtual polycrystals \cite{frankel2020prediction, mianroodi2021teaching,pandey2021machine}. In these efforts, polycrystals are represented by a grid of voxels containing microstructural descriptors such as lattice orientation, and the CNN learns spatial correlations between the voxels to generate full-field predictions. In particular, CNNs take advantage of the grid structure of the data to learn filter structures consisting of series of weights. These weights group (pool) together neighboring grid values (in this case microstructural features) to capture and predict the effects of neighborhood. The results of this approach are very promising, but an issue is that (most) materials are not naturally structured in a grid-like fashion (although due to data collection strategies, materials are often represented as such). In the cases of polycrystalline materials at the microscale, the grains themselves are more naturally represented in an unstructured, connected format often referred to as a graph. 

Broadly, a graph is a structure in which `vertices' or `nodes' are connected through `edges'. Both nodes and edges can further be described via associated attributes or features. In the case of a polycrystal represented via a graph, grains are considered as nodes, while grain boundaries are considered as edges. Features of nodes (grains) could include local micromechanical properties such as stiffness or strength, as well as microstructural features such as lattice orientation or dislocation density. Features of edges (grain connections or boundaries) may be distance between grains or grain boundary characteristics. GNNs, which are the focus of this work, adapt the data pooling technique of CNNs on structured data to the pooling of neighboring data on unstructured graph data. Instead of pooling features from neighboring grid points, features are pooled from connected nodes~\cite{hamilton2017inductive}. A recent study has demonstrated the utility of GNNs for predicting magnetorestriction response in a polycrystalline material \cite{dai2021graph}. 

In this study, we apply GNNs to predict the elastic response in two alloy systems: Low Solvus High-Refractory (LSHR) Ni superalloy and Ti 7wt\%Al (Ti-7Al), representing cubic and hexagonal elastic responses, respectively. Two GNN surrogate models for predicting LSHR and Ti-7Al grain-averaged elastic response (grain stress tensor components along the loading direction) that utilize Gaussian Mixture convolutions are implemented, trained, and preliminarily tested using CEFEM simulations that explicitly model grain microstructure. During training, the accuracy of the predictions are compared against predictions from traditional mean-field theories and reserved CEFEM simulations. The GNN models trained with CEFEM simulations are then transferred to a new data domain to predict the grain-averaged elastic response in polycrystals with microstructures measured with near-field HEDM. GNN model predictions are then compared to stresses measured using far-field HEDM. As part of the GNN model training effort, learning rates of the models and accuracy of using various nodal features for stress predictions are explored. 

In this paper, vectors are generally lower-case bold characters ($\bm{a}$), second order tensors are upper-case bold characters ($\bm{A}$), and fourth order tensors are underlined, bold characters ($\underline{\bm{A}}$). Unless otherwise noted, quantities are expressed in the sample frame. Prime characters are generally used to indicate quantities in subsequent layers within GNNs, i.e., $a'$ for Hidden Layer n+1, while an overbar $\bar{a}$ indicates an average.
\section{Methods}

\begin{figure}[h]
	\centering
	\includegraphics[width=0.9\textwidth]{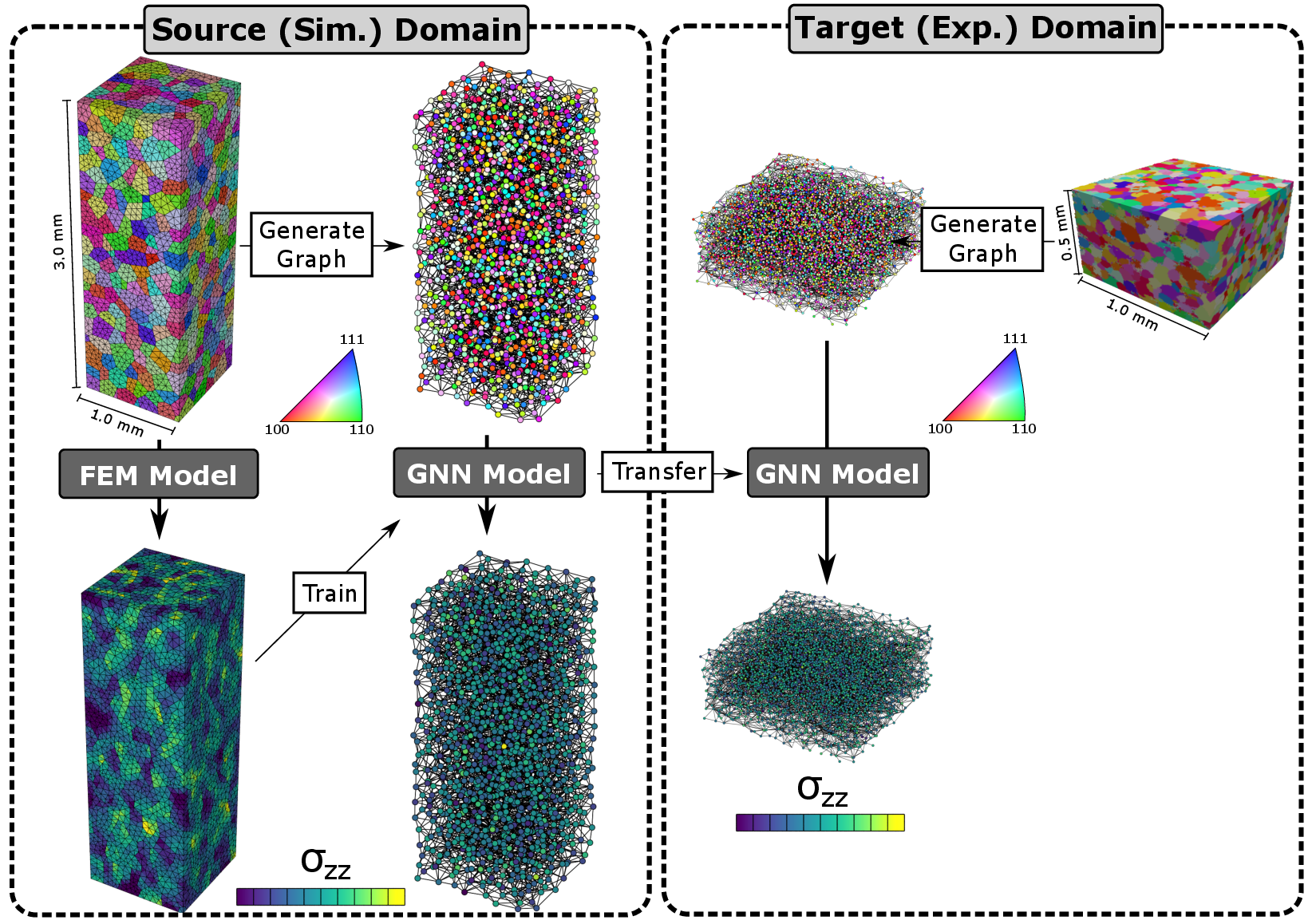}
	  \caption{Connectivity of the various efforts used to evaluate the accuracy of GNN surrogate models for predicting the stress response in individual grains during elastic loading.}
	  \label{fig:overview}
\end{figure}

In this section, we give a broad overview of the various methods employed in this work for GNN training and accuracy evaluation including: virtual sample generation, CEFEM modeling, HEDM, and GNN surrogate modeling. In addition, details regarding the various training and testing data used are provided. A schematic of the various components of the effort, displaying LSHR data, are given in Fig. \ref{fig:overview}. Briefly, a transfer learning approach \cite{weiss2016survey} is taken in which the GNN models for LSHR and Ti-7Al are trained using simulated data from microscale CEFEM modeling (the Source Domain) and then transferred to predict the mechanical response in a microstructure measured experimentally via HEDM (the Target Domain). 

\subsection{Virtual Sample Generation}
\label{sec:neper}

Virtual polycrystals upon which mechanical deformations  are imposed and used to create microstructure graphs are generated using the Neper polycrystal generation and meshing software~\cite{neper_paper,renversade,neper_website}. Generally, Neper utilizes Laguerre tessellations to generate polycrystalline samples. Laguerre tessellations produce convex, space-filling grains in a user-specified sample domain. Specifically, Neper allows for user-specified target distributions of grain size and grain shape. This affords the ability to create a wide range of microstructures with various geometric features. An attendant finite element mesh is then generated via Neper using Gmsh \cite{geuzaine2009gmsh}, in which the geometric features of the microstructure (tessellation) are preserved. 

Here, Neper is used to generate 50 virtual polycrystals (25 LSHR and 25 Ti-7Al) for elastic finite element simulations. The simulated domains are \SI{1}{\milli\meter} $\times$ \SI{1}{\milli\meter} $\times$ \SI{3}{\milli\meter}, each containing 1,500 grains. Each polcyrystal had approximately 16,500 shared grain boundaries which serve as edges. Grain size and shape distributions are set to create nominally equiaxed grains with diameters of approximately \SI{150}{\micro\meter} and minimal spread. Each polycrystal is meshed with ten node tetrahedral elements and approximately 120,000 elements per sample. Orientations are assigned randomly to grains from the cubic and hexagonal fundamental regions for the LSHR and Ti-7Al virtual specimens respectively. 

\subsection{Polycrystal Anisotropic Elasticity Data}
\label{sec:fepx}

For generating training data for the GNN surrogate elasticity model deformation response, we utilize the finite element solver within FEPX~\cite{fepx_arxiv,fepx_website} which interfaces directly with tessellations and meshes generated via Neper \cite{quey2022neper}. Generally, FEPX considers the elasto-viscoplastic deformation response of single crystals (grains) belonging to explicit representations of polycrystalline aggregates. However, here, viscoplasticity is inhibited by ensuring grain level stresses are significantly lower than the yield point. In the model, grain-to-grain interactions are assumed to be rigid (no grain boundary sliding or separation). As the implementation and use of these models are well established~\cite{Asaro1985,marin_a,marin_b,roters}, and as this study considers elastic deformation, only a truncated description of the model as implemented in FEPX is included below. Please refer to~\cite{fepx_arxiv} for a complete description of kinematics, models, and finite element implementation. For the uniaxial deformation studied in this work, loading is along the $\bm{z}$ direction in the sample frame.

At each point in a finite element mesh, FEPX considers the elastic response to be governed by the anisotropic form of Hooke's law:
\begin{equation}
    \label{eq:hooke}
    \bm{\sigma}=\underline{\bm{C}}(\bm{r}):\bm{\varepsilon}
    \quad ,
\end{equation}
where $\bm{\sigma}$ is the stress, $\underline{\bm{C}}$ is the lattice-orientation-dependent elastic stiffness tensor, $\bm{r}$ is the lattice orientation (a coordinate transformation from crystal frame to sample frame) of a given crystal, and $\bm{\varepsilon}$ is the strain, assumed to be fully elastic. Due to symmetry, crystals with cubic symmetry have three independent constants ($C^C_{11}$ , $C^C_{12}$ , and $C^C_{44}$  in Voigt notation in the crystal frame), while crystals with hexagonal symmetry have five independent constants ($C^C_{11}$ , $C^C_{12}$ , $C^C_{13}$ , $C^C_{33}$ , and $C^C_{44}$  in Voigt notation in the crystal frame). We note that due to the deformation decomposition formulation in FEPX, in hexagonal crystals further dependence is required in the form of $C^C_{33} = C^C_{11}+C^C_{12}-C^C_{13}$. The elastic moduli employed for LSHR~\cite{turner2012two} in this work are: $C^C_{11}=247$, $C^C_{12}=147$, $C^C_{44}=$ \SI{125}{\giga\pascal}, while the Ti-7Al~\cite{Fisher1964} moduli are: $C^C_{11}=162$ , $C^C_{12}=92$ , $C^C_{13}=69$ , $C^C_{33}=185$ , and $C^C_{44}=$ \SI{45}{\giga\pascal}.

In uniaxial deformation, the effective stiffness along the loading direction can be approximated using a directional modulus $E(\bm{r})$. Generally, in grains embedded in polycrystals, the stress along the loading direction is correlated to the directional modulus \cite{kocks1998texture}. Under uniaxial stress along the $\bm{z}$ in the sample frame, the directional modulus is defined as:
\begin{equation}
    \label{eq:dirmod}
    E(\bm{r})=\frac{\sigma_{zz}}{\varepsilon_{zz}(\bm{r})}
    \quad .
\end{equation}
To further describe the anisotropy of the grains embedded within a polycrystal during uniaxial deformation, we define effective transverse contraction ratios $\nu_x(\bm{r})$ and $\nu_y(\bm{r})$:
\begin{equation}
    \nu_x(\bm{r})=\frac{\varepsilon_{xx}(\bm{r})}{\varepsilon_{zz}(\bm{r})}
    \quad ,
\end{equation}
and 
\begin{equation}
    \nu_y(\bm{r})=\frac{\varepsilon_{yy}(\bm{r})}{\varepsilon_{zz}(\bm{r})}
    \quad .
\end{equation}

Each of the 50 virtual polycrystals generated (see \S \ref{sec:neper}) are deformed elastically in FEPX. As all uniaxial loading linear elasticity solutions can simply be scaled to account for increasing or decreasing deformation, each polycrystal is only deformed with a single deformation increment to an applied strain of 0.1\% (chosen arbitrarily). Minimal displacement boundary conditions are employed on the top and bottom surfaces of the specimen to prevent rigid translation and rotation of the virtual specimens without impeding contraction of the specimen perpendicular to the loading direction.

\subsection{High-Energy X-ray Diffraction Microscopy Data}
\label{sec:hedm}

For final GNN evaluation, graph data derived from experimentally measured microstructures and micromechanical response measured using HEDM are utilized. HEDM is comprised of two variants, near-field and far-field, capable of non-destructively characterizing the microstructure and micromechanical response of polycrystalline materials \cite{poulsen2004three,bernier2011far,nygren2020algorithm}. The commonality between the two techniques is the utilization of forward projection simulations of X-ray diffraction peaks of individual grains to reconstruct information about the local lattice state of crystalline materials in 3D. LSHR and Ti-7Al samples were probed using the near-field variant to measure the grain structure, orientation, and connectivity of the grains, and the far-field variant was used to determine the stress in the same grains during elastic loading. Detailed descriptions of the data collection for the LSHR sample can be found in \cite{musinski2021statistical} and for the Ti-7Al sample in \cite{pagan2021analysis}, but summaries of the methodology and specimens are given below.

The near-field variant (nf-HEDM) is capable of reconstructing a 3D voxelized distribution of lattice orientation with resolution on the order of \SI{}{\micro\meter} from series of diffraction images collected as a specimen is rotated. In the near-field variant, the detector is placed approximately \SIrange{5}{10}{\milli\meter} away from the specimen, making the measurements sensitive to the spatial locations of diffracting volumes  \cite{pagan2014connecting}. For this work, the LSHR and Ti-7Al microstructures were reconstructed with \SI{5}{\micro\meter} voxel spacing. The reconstructed volume for the LHSR specimen was \SI{1.0}{\milli\meter} $\times$ \SI{1.0}{\milli\meter} $\times$ \SI{0.5}{\milli\meter} with the short direction aligned with the loading direction. Similarly, the reconstructed volume for the Ti-7Al specimen was \SI{1.0}{\milli\meter} $\times$ \SI{1.0}{\milli\meter} $\times$ \SI{0.5}{\milli\meter}, again with the short direction along the loading direction.

The far-field variant (ff-HEDM) can reconstruct the average grain orientation, position, and elastic strain state of grains embedded in a polycrystal during \emph{in situ} loading \cite{miller2020understanding}. In this HEDM variant, a large area detector is placed approximately \SI{1}{\meter} away from the specimen. This positioning provides more sensitivity to peak shifts due to changes in orientation and strain state, as opposed to locations of diffraction events as utilized for nf-HEDM reconstructions. Here ff-HEDM measurements collected prior to and during \emph{in situ} elastic loading to applied strains near 0.001 --- 0.0015 and 0.0009 for the LSHR and Ti-7Al respectively --- are used for GNN prediction evaluation. The full elastic strain tensors from individual grains are then used to calculate the stresses in each of the grains, including the component of stress along the loading direction which is of interest here. The uncertainty per elastic strain component for HEDM measurements are generally reported to be $10^{-4}$ \cite{hurley2018characterization}.

\subsection{Graph Neural Network Modeling}

As previously described, the surrogate models employed in this work are graph
neural networks (GNNs). The GNNs are comprised of layered graphs in which the
features of each node in subsequent graph layers are weighted combinations
(convolutions) of $N_F$ nodal features from neighbors in the previous graph
layer. The weights may be independent or functions of nodal or edge features and
are learned throughout the training process. Similar to other neural network
formulations, nonlinear activation functions (functions which control the flow of
information between layers, i.e., nominally turning on and off) control the
passing of nodal feature information between layered graphs.  The input data is
the graph itself --- a set of nodes and a set of edges --- along with
features associated with the nodes and edges expected to be useful for
predicting stress. Here graphs are generated from CEFEM or HEDM data using a series of custom Python scripts.

Nodal features represent different microstructural descriptors or local mechanical properties. Various nodal
features, and combinations of nodal features, were explored, including the
directional modulus ($E(\bm{r})$), contraction ratios
($\nu_x(\bm{r})$,$\nu_y(\bm{r})$), volume ($V$), and lattice orientation
($\bm{r}$, Rodrigues parameterization) of each grain.  We note that no appreciable accuracy difference was observed utilizing different orientation parameterizations (e.g., Euler angles or other angle-axis parameterizations). These nodal features alone could be useful for predicting stress.  However,
rather than just treating all nodes independently, the GNN combines these
features with the same features of connected nodes in the graph, along with
additional edge features associated with the connections. The edge features used here
for a connected pair of grains $i$ and $l$ are the coordinates of the vector between the grain
centroids $\bm{p}_i$ and $\bm{p}_{l}$:
\begin{equation}
\bm{e}_{il}=\bm{p}_{i}-\bm{p}_{l}
\quad .
\label{eq:ef}
\end{equation}

\begin{figure}[h]
	\centering
	\includegraphics[width=1.0\textwidth]{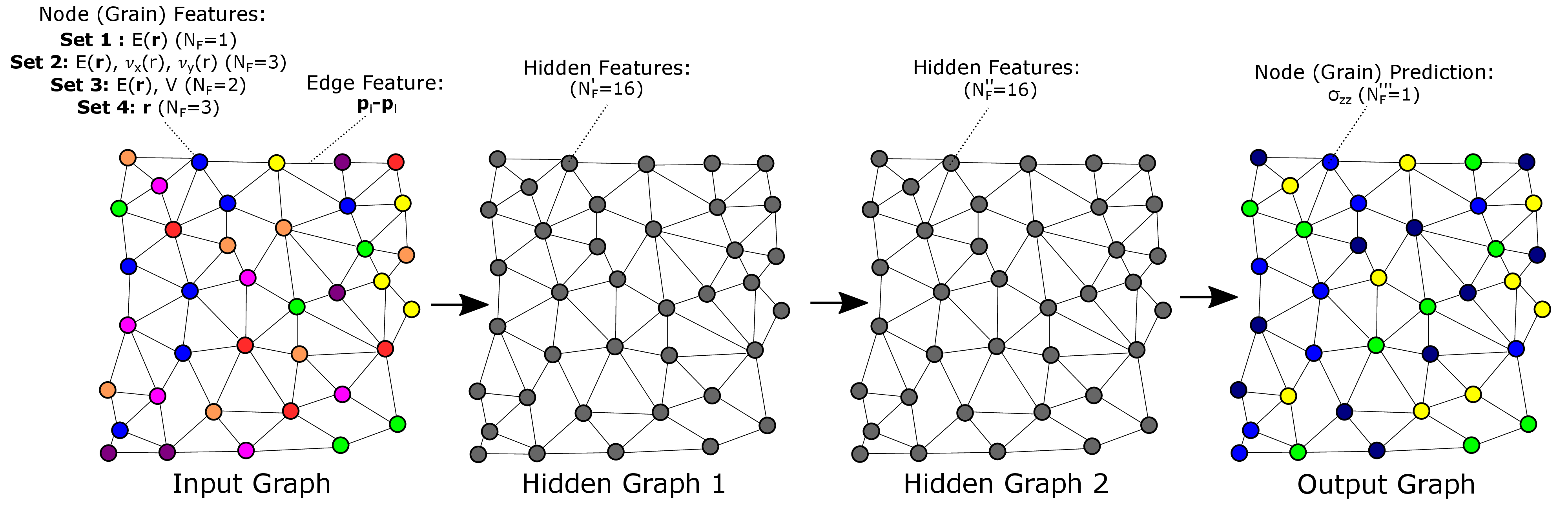}
	  \caption{Schematic of the GNN surrogate architecture used for this work. An input graph with various sets of nodal features maps to a final output layer for predicting stress along the loading direction $\sigma_{zz}$}
	  \label{fig:arch}
\end{figure}

Rather than just looking at immediate connections, the GNN also works
recursively, learning a vector representation of node features at subsequent
layers. For this work, a common GNN architecture was utilized with two hidden graph layers. A schematic of the architecture is given in Fig. \ref{fig:arch}. 
Each node within the two hidden graph layers contains 16 features obtained from a convolution of the previous layer's features over neighboring nodes. For the first hidden graph layer, the convolution is over the input nodal features. Two hidden graph layers provide substantial flexibility for learning the non-linear relationship between grain lattice orientation and
elastic response. After the two hidden graph layers, the second 16-dimensional vector representation of
the nodal features is passed through a final dense layer that maps to a single
nodal feature, which is the predicted stress tensor component $\sigma_{zz}$.  In total, there are
four layers: one input graph with the input nodal features, two hidden
convolutional graphs with learned features, and the final mapping layer. While the number of nodal features in the two hidden graph layers could be
slightly oversized for the problem at hand, over-fitting was only observed when
very small amounts of training data were employed, as will be seen.

The graph convolution operator used in the anisotropic elasticity surrogate models is a Gaussian
Mixture Model \cite{monti2017geometric}, implemented in PyTorch Geometric
\cite{fey2019fast}, with the form (no sum over indices $i$ and $j$):
\begin{equation}
x'_{ij}=\frac{1}{N_N}\sum_{k=1}^{N_F} \theta_{jk} \sum_{l \in N(i)} w_k(\bm{e}_{il}) x_{lk}
\quad ,
\end{equation}
where $x$ is a nodal feature (either the initial inputs, or the learned vector
representation in the hidden layers), $i$ indicates node (grain) number ranging
from 1 to $N_G$, $k$ indicates a nodal feature in a current layer ranging from 1
to $N_F$, $j$ indicates a nodal feature in a subsequent layer ranging from 1 to
$N_F'$, $w$ is a Gaussian weighting function, $l$ indexes a neighbor to grain
$i$, $N(i)$ is the set of indices of nodes connected to
$i$, $\bm{e}$ is the vector of edge features between nodes $i$ and $l$ from
Eq.~\ref{eq:ef}, and $\theta$ is a (learned) weight for a given nodal feature in
the current layer. The weighting functions are comprised of $N_K$ Gaussian kernels:
\begin{equation}
 w_k(\bm{e}_{il})=\sum_{M=1}^{N_K}\exp\left(-\frac{1}{2}(\bm{e}_{il}-\bm{\mu}_{m})^T\bm{\Sigma_m}(\bm{e}_{il}-\bm{\mu}_{m})\right)
 \quad ,
 \label{eq:gauss}
\end{equation}
with $\bm{\mu}_m$ and $\bm{\Sigma_m}$ being Gaussian kernel and diagonal
covariance matrices. In total, $\theta$, $\bm{\mu}$, and $\bm{\Sigma}$ are
coefficients learned during the training process. Using the 3D vector between
grain centroid as edge features (Eq. \ref{eq:ef}), each layer in the neural
network requires $N'_F \times (N'_F \times ( 1 + N_K \times (3 + 3)))$
coefficients to be learned. The choice of a Gaussian Mixture convolution (GM) was informed by our physical understanding of deformation compatibility and mechanical
equilibrium. In this model, GNN has sufficient degrees of freedom
to weight various positions around a grain differently (e.g., parallel and
transverse to applied uniaxial load). In addition, during the process of developing the GNNs presented in this work, several spectral- and spatial-based convolution operators were initially tested. In general, spatial-based convolutions performed better than spectral-based, consistent with our understanding of the spatial nature of stress equilibrium. The GM convolution operator presented here was the best performing, while many other common convolutional operators (such as that found here in \cite{kipf2016semi}) did not outperform the accuracy of more traditional mean-field theories, and as such, are not presented.

For final fitting, seven Gaussian kernels ($N_K=7$) were used for the Gaussian convolution
operator in Eq.~\ref{eq:gauss} with the goal of providing the surrogate model
the freedom to weight neighborhood grains arranged in various positions
differently (i.e., parallel and transverse to loading). Leaky ReLU was chosen
for the activation function, with a scaling value of -0.1 for input values less
than 0 as it provided better accuracy than a standard ReLU activation function. Training of the surrogate models was performed by minimizing the mean
square error (MSE) between grain average stress components in the training data
and those predicted by the GNN surrogate model using Adam stochastic gradient descent
with a learning rate parameter of 0.01. Training of a surrogate model for 10,000 epochs nominally
takes 2-15 minutes depending on the amount of training data (1,500 grains to 3,000 grains) using an NVIDIA
Quaddro GPU with 5 GB of memory.

\section{Results}
\label{sec:results}

The results are divided into three subsections. The first subsection covers the learning behavior of GNN surrogate models using CEFEM data. The second subsection details how the trained GNN surrogate models perform predicting micromechanical response in comparison to reserved CEFEM data is described. The final subsection analyzes the performance of the trained surrogate GNN models in predicting mechanical response of microstructures characterized experimentally.

\subsection{Model Training}

After generating the various graph data from both CEFEM simulations and HEDM results (Sec \ref{sec:fepx} and \ref{sec:hedm}) for GNN surrogate model training, a study was completed to examine the performance of the GNNs predicting the stress response along the loading direction ($\sigma_{zz}$) in individual grains. For GNN model training, graphs generated from various numbers of CEFEM simulations (1, 5, 10, and 20 corresponding to 1,500, 7,500, 15,000, and 30,000 grains) were used to train the LSHR and Ti-7Al GNN surrogates. Respectively for each training scenario, 1, 2, 3, and 4 CEFEM simulations (1,500, 3,000, 4,500, and 6,000 grains) were reserved for evaluating the accuracy of the surrogate model predictions. Accuracy of surrogacy models in predicting simulated stresses were quantified using the mean of a 1-Norm error $e$:
\begin{equation}
\bar{e}(\sigma_{zz})=\frac{1}{N_G} \sum \frac{|\sigma_{zz,i}^{\mathrm{SIM}}-\sigma_{zz,i}^{\mathrm{GNN}}|}{\sigma_{zz,i}^{\mathrm{SIM}}}
\quad.
\end{equation}
During training, this error metric provides a direct comparison of the accuracy of GNN predictions versus the full-field CEFEM simulations.

\begin{figure}[h]
	\centering
	\includegraphics[width=0.9\textwidth]{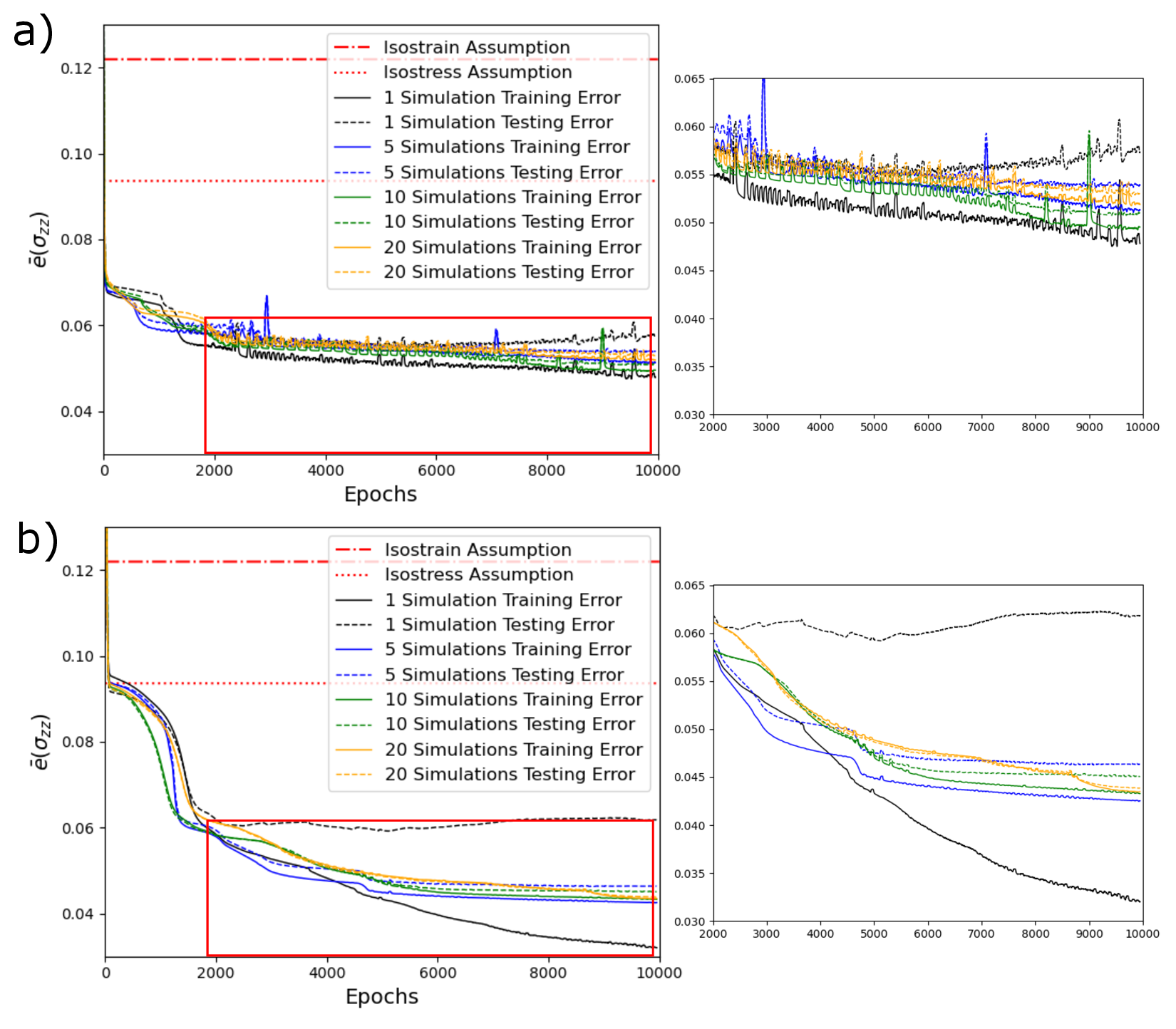}
	  \caption{Learning rates of LSHR grain-scale elastic response using various numbers of CEFEM simulations for training and testing. Insets show a magnified view of the learning rate at larger epochs. a) Learning rates using the directional modulus $E(\bm{r})$ as a nodal feature. b) Learning rates using the lattice orientation (components of the Rodrigues vector $\bm{r}$) as nodal features.}
	  \label{fig:lshr_learning}
\end{figure}

Fig. \ref{fig:lshr_learning} shows the learning rates (epoch vs error) for LSHR elastic response using the various numbers of training data sets described above. Fig. \ref{fig:lshr_learning}a shows the learning rate using the directional modulus $E(\bm{r})$ as a nodal feature, while Fig. \ref{fig:lshr_learning}b shows the learning rates using components of the Rodrigues vector $\bm{r}$ describing a grain's lattice orientation as nodal features. It is important to note that while these quantities are related (the directional modulus is a function of the grain lattice orientation), the directional modulus is a micromechanical property that should have a linear relationship to the stress state, while the components of the lattice orientation are a microstructural feature that should have a non-linear relationship to the stress state. Solid lines show the mean error $\bar{e}$ of grain stress predictions of the surrogate model on the training data while the dashed lines show the mean error in comparison to the reserved testing data sets. As a benchmark, mean errors as predicted by various mean-field theories (models), isostress (all grains have the same stress state as the macroscopic stress, $\sigma_{zz}=\tilde{\sigma}_{zz}$) and isostrain (all grains have the same strain state as macroscopic strain, $\sigma_{zz}=E(\bm{r})\tilde{\varepsilon}_{zz}$), are provided. In most cases, there is no evidence of over-fitting which would be seen as divergence between mean errors on predictions of training and reserved testing data, except for the case when only one training data set is used with lattice orientation used as the nodal feature input. The `spikes' during the learning process are related to the use of Adam stochastic gradient descent algorithm for fitting, which will perturb the solution intermittently to try to ensure that a global minimum is reached. 

In Fig. \ref{fig:lshr_learning}, we can see that using the GNN surrogate model with either the directional modulus or lattice orientation as nodal features outperforms the predictions of the mean-field theories (red dashed and dotted lines) reaching average errors of approximately 0.05 (5\%) for LSHR, with the lattice orientation performing slightly better after 10,000 epochs of training. The fact that surrogate models using directional modulus and lattice orientation have approximately the same performance indicate that nominally the same information is encoded into the two grain descriptors. This can be rationalized in that the directional modulus is calculated using the lattice orientation and the single crystal elastic moduli as previously mentioned. However, this relationship between lattice orientation and directional stiffness is non-linear and naturally the rate at which the relationship is learned is slower, but the GNN surrogate model does in fact learn the relationship. We note that when using only a single hidden layer, the surrogate model is incapable of learning this relationship (not shown). 

\begin{figure}[h]
	\centering
	\includegraphics[width=0.9\textwidth]{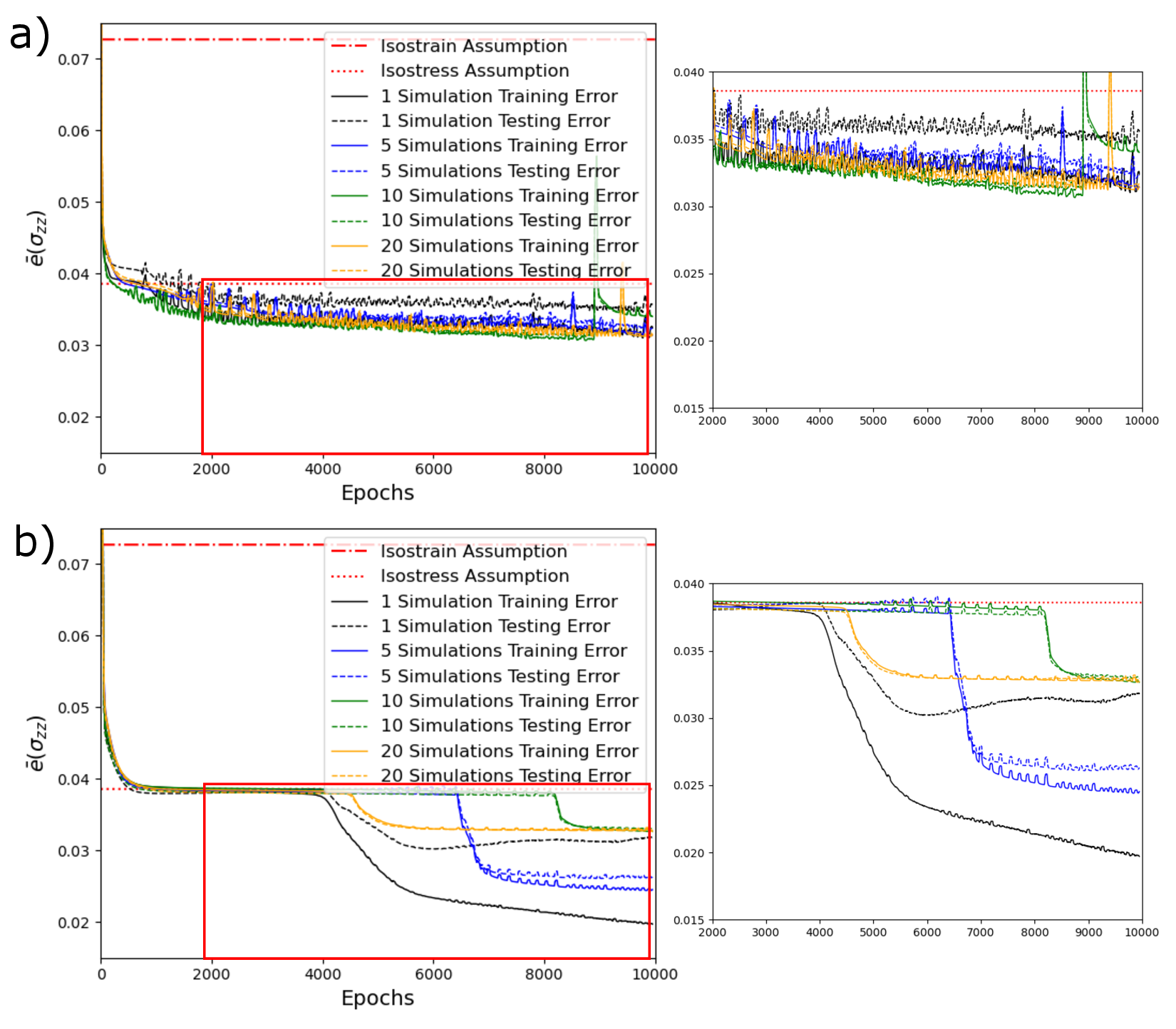}
	  \caption{Learning rates of Ti-7Al grain-scale elastic response using various numbers of CEFEM simulations for training and testing. Insets show a magnified view of the learning rate at larger epochs. a) Learning rates using the directional modulus $E(\bm{r})$ as a nodal feature. b) Learning rates using the lattice orientation (components of the Rodrigues vector $\bm{r}$) as nodal features.}
	  \label{fig:ti7_learning}
\end{figure}

Similar to Fig. \ref{fig:lshr_learning}, Fig. \ref{fig:ti7_learning} shows learning rates of the grain-scale elastic response using different numbers of CEFEM simulations for training and using directional modulus (Fig. \ref{fig:ti7_learning}a) and lattice orientation (Fig. \ref{fig:ti7_learning}b) as input nodal features for Ti-7Al. In Fig. \ref{fig:ti7_learning}b, it can be seen that using 1 and 5 CEFEM simulations for training along with lattice orientation as nodal features show signs of over-fitting for Ti-7Al as opposed to only 1 FEM simulation for the LSHR data. Like the LSHR, using directional modulus or lattice orientation as input nodal features gives approximately the same performance  (0.035 mean error) at large epochs. The magnitude of the final mean error is lower than that of the LSHR (0.035 for Ti-7Al versus  0.05 for LSHR) which is discussed further in \S \ref{sec:anisotropy}. In addition, the GNN surrogate model only performs slightly better than using an isostress assumption to determine grain average stresses in the Ti-7Al which will be discussed.

\subsection{GNN Surrogate Model Performance: Source Domain (CEFEM Data)}

\begin{figure}[h]
	\centering
	\includegraphics[width=0.9\textwidth]{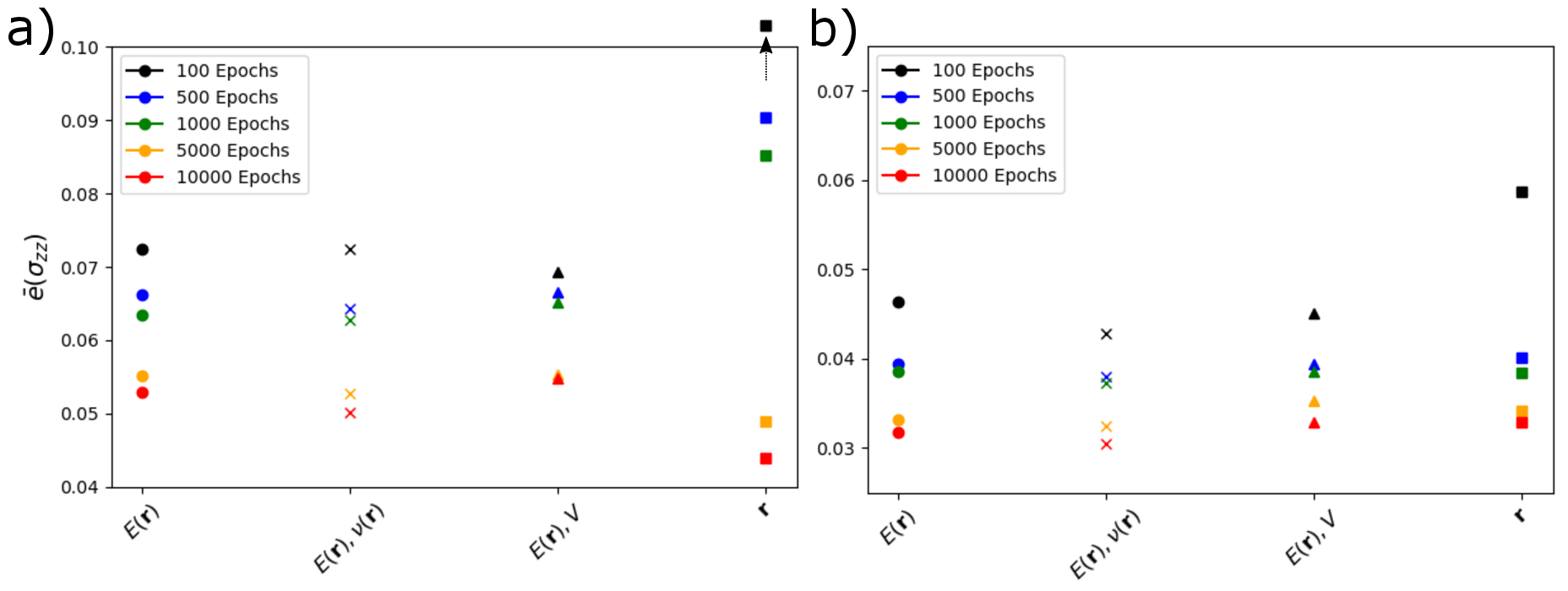}
	  \caption{Summary of learning rates for various nodal features included into the GNN surrogate models for a) LSHR and b) Ti-7Al.}
	  \label{fig:learning_summary}
\end{figure}

Besides directional modulus and lattice orientation, the performance of GNN surrogate models in predicting stresses in the Source Domain (i.e., domain where the surrogate models are trained using CEFEM data) using other sets of nodal features were also tested. These include incorporating transverse contraction ratios $\nu_x(\bm{r})$ and $\nu_y(\bm{r})$ in addition to the volume $V$ of grains into the fitting, for a total of four different nodal feature sets. A summary of the training results for the different nodal feature sets using 20 CEFEM simulations (30,000 grains) for training and 4 CEFEM simulations for testing (6,000 grains) are shown in Fig. \ref{fig:learning_summary}. This figure compares the results of the GNN based stress predictions against those from the full-field CEFEM simulations reserved solely for testing and not included in the model training, providing a bench-mark against existing modeling capabilities. Beyond what can be seen in Fig. \ref{fig:lshr_learning} and Fig. \ref{fig:ti7_learning}, the inclusion of transverse contraction ratios generally does not improve accuracy. This is consistent with the observation that the directional modulus and lattice orientation nominally have nominally the same accuracy: if transverse contraction ratios improved accuracy, it would be expected that lattice orientation would also perform better (contraction ratios are calculated from orientation). Somewhat surprisingly, inclusion of grain volume $V$ into the nodal features does not improve the surrogate model prediction accuracy. For this work, however, spread of grain size is minimal, and we note that this observation may not be general. Volume may be a necessary descriptor for accuracy in other microstructures with larger or bi-modal distributions of grain size.

\begin{figure}[h]
	\centering
	\includegraphics[width=0.9\textwidth]{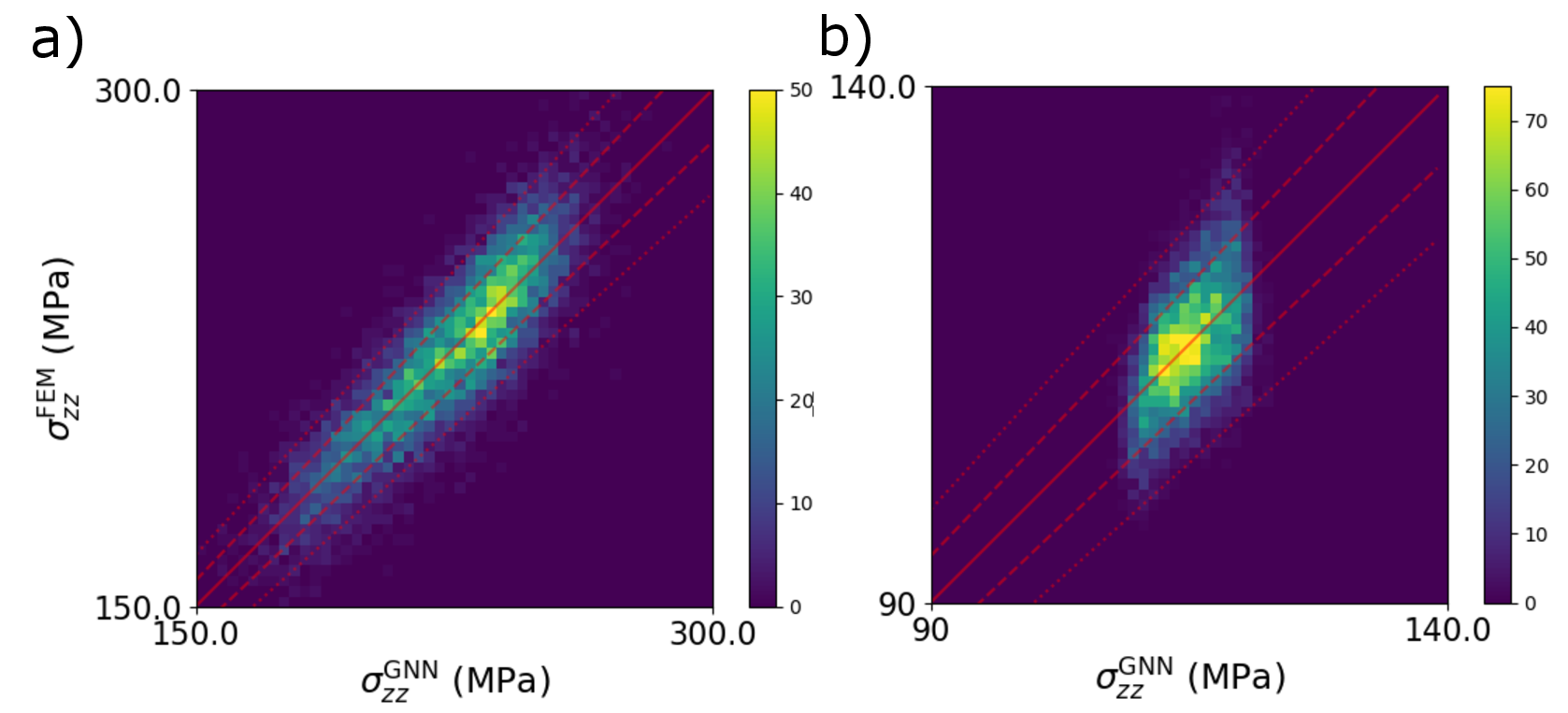}
	  \caption{Comparisons of stress predictions along the loading direction ($\sigma_{zz}$) from GNN surrogate models using lattice orientation as nodal features compared to CEFEM simulations for  a) LSHR and b) Ti-7Al.}
	  \label{fig:fem_fit}
\end{figure}

To examine if there were systematic errors in the GNN surrogate model predictions, the stress predictions from the surrogate models using lattice orientation (Rodrigues vectors) as nodal features were compared to stresses from the CEFEM reserved testing data sets. Comparison of surrogate model predictions using 20 CEFEM simulations for training and compared to stresses from 4 testing CEFEM simulations are shown in 2D histograms in Fig. \ref{fig:fem_fit}a and Fig. \ref{fig:fem_fit}b for LSHR and Ti-7Al respectively. The solid red diagonal lines correspond to perfect correspondence (100\% accuracy) while the dashed and dotted lines correspond to 5\% and 10\% error bounds respectively. For the LSHR, 64\% of the GNN stress predictions fall within the 5\% bound, and 93\% of stress predictions fall within the 10\% bound. It should be noted that spread of stresses across the LSHR grains is significantly larger than that seen in the Ti-7Al due to the much larger elastic anisotropy. The LSHR fit shows no evidence of systematic bias and is able to predict stresses within the full ranges of stresses generated by the significant elastic anisotropy in LSHR. For the Ti-7Al, the GNN surrogate model captures most of the grain stress spread, but appears to be truncating the bounds of the most extreme values of the prediction. In Fig. \ref{fig:fem_fit}b, 78\% and 98\% of the GNN stress predictions fall within the 5\% bound and 10\% bound respectively. These observations will be discussed further in \S \ref{sec:anisotropy}.



\subsection{GNN Surrogate Model Performance: Target Domain (HEDM Data)}

\begin{figure}[h]
	\centering
	\includegraphics[width=0.9\textwidth]{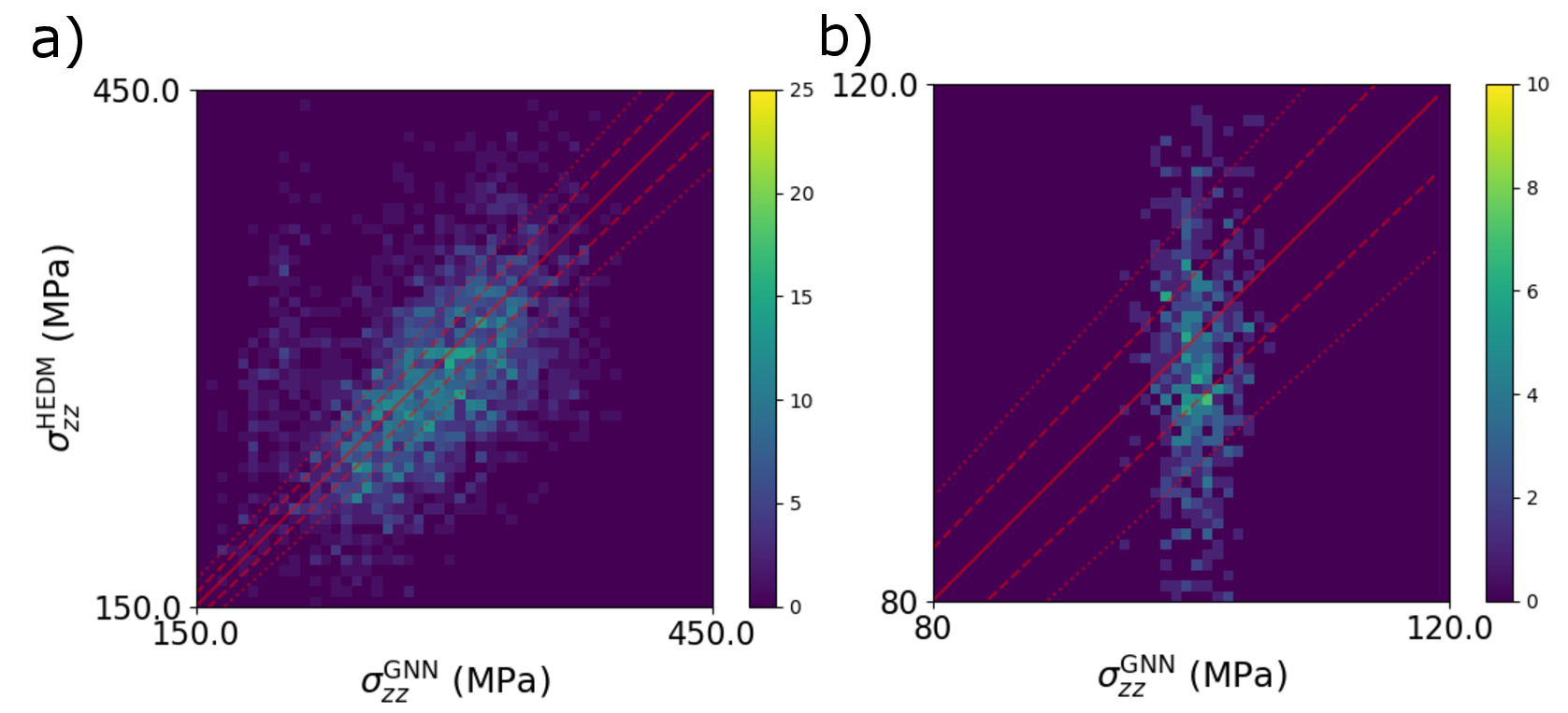}
	  \caption{Comparisons of stress predictions along the loading direction ($\sigma_{zz}$) from GNN surrogate models using lattice orientation as nodal features compared to those measured using ff-HEDM during in situ elastic loading for  a) LSHR and b) Ti-7Al.}
	  \label{fig:hedm_fit}
\end{figure}

After evaluating the performance of the surrogate models in the Source Domain, the surrogate models were then transferred to the Target Domain where stresses during elastic loading was predicted from a graph generated from a 3D microstructure measured using nf-HEDM. As the performance of various nodal features used for prediction were fairly similar (Fig. \ref{fig:learning_summary}), only GNN surrogate models using lattice orientation as nodal features were tested as lattice orientation is directly measured using HEDM. The stresses predicted by the GNN surrogate models were then compared to the stresses measured from the same microstructure during elastic loading using ff-HEDM. Comparison of the GNN predicted versus ff-HEDM measured stresses for the LSHR  and Ti-7Al specimens are shown in Figs. \ref{fig:hedm_fit}a and \ref{fig:hedm_fit}b, respectively. For comparison, the stresses for 2,896 LSHR grains and 592 Ti-7Al grains were used. For prediction of stresses using the GNN surrogates, the output stresses are scaled by the ratio of macroscopic applied strain in the experiments (0.0015 for the LSHR and 0.0009 for the Ti-7Al) to the macroscopic applied strain for the CEFEM simulations in the Source Domain (0.001). As in the Source Domain, there is no systematic error in the prediction of stresses for LSHR observed in Fig. \ref{fig:hedm_fit}a. However, the mean of the errors are higher ($\bar{e}=13.3\%$), with 27\% (783) of grains having errors less than 5\% and 49\% (1,428) of grain having errors less than 10\%. There are multiple possible contributions to this elevated error including measurement error, non-perfect application of uniaxial stress (e.g., small bending or torsion stresses), or more complex grain geometries and interactions than those created using Laguerre tessellations for the simulations. In Fig. \ref{fig:hedm_fit}b, we can see the GNN stress predictions are fairly tightly clustered around the average stress in the specimen (100 MPa). There appears to be a systematic error in which the GNN predictions for Ti-7Al do not capture the full spread of grain stresses in the simulation, similar to observations in the Source Domain, but increased due to the various factors likely contributing to errors with use of experimental data. Although the systematic error is present, the Ti-7Al stress predictions still have a lower mean error ($\bar{e}=9.2\%$) and higher percentage of grains within the 5\% bound (41\%, 227 grains) and 10\% bound (71\%, 392 grains) in comparison to the LSHR due to the lower elastic anisotropy of the Ti-7Al. 

\section{Discussion}

Here the efficacy of using GNN surrogate models to predict grain-scale elastic response on both simulated and experimentally measured microstructures (represented as graphs) was explored. The GNN models use a graph convolution (Gaussian Mixture) to capture the inherent non-locality in both mechanical equilibrium and deformation compatibility which govern mechanical response. Elasticity was chosen as an initial test case for exploring prediction of micromechanical response due to our generally strong understanding of the constitutive relationship (Hooke's Law) and primary material characteristics (the single crystal elastic moduli and lattice orientation of a grain) governing elasticity at the grain-scale. It was found that, once trained, the accuracy of the GNN surrogate models exceeded that of both isostrain (Taylor) and isostress (Sachs) mean-field theory assumptions in both a highly anisotropic FCC alloy (LSHR) and a moderately isotropic HCP alloy (Ti-7Al). Various sets of microstructural and micromechanical descriptors were tested and found to have generally similar accuracy after full training (Fig. \ref{fig:learning_summary}), although it is noted that the choice of descriptors was informed by our generally strong understanding of the elastic response. Importantly though, the two hidden layer architecture employed was demonstrated to be capable of learning the complex non-linear relationship between lattice orientation and directional stiffness as evidenced by the similar final performances.

In a novel approach, transfer learning was successful in enabling the prediction of the mechanical response of microstructures (as represented by graphs) that were experimentally measured using HEDM, as opposed to only predicting the response of synthetic microstructures. The ability to move between various data domains opens up the possibility of training schemes that could further improve the accuracy of the model. In particular, using both simulated and experimental data to train GNN surrogate models is a promising avenue for future efforts as both data types have strengths and weaknesses for use in learning material response. Large amounts of simulated data is generally more readily acquired than experimental data, however simulations are generally not capable of capturing phenomena not included into the model. Experimental data will naturally represent the physics of the phenomenon at hand, but is susceptible to measurement error and bias. Creating training data sets in the future containing both simulation and experimental data can help to further improve the accuracy of GNN surrogate model predictions.

\subsection{Mechanical Anisotropy}
\label{sec:anisotropy}

At this point, it is worth discussing differences between the elastic responses of LSHR and Ti-7Al and the utility of GNN surrogate models for predicting elastic responses. While the LSHR is crystallographically more symmetric, it is actually significantly more elastically anisotropic. The directional moduli in the LSHR vary by a factor of approximately 2.5 (ratio of maximum to minimum), while the directional moduli of the Ti-7Al only vary by a factor of approximately 1.5. The value of the use of the GNN surrogate model in comparison to mean-field theory will increase as a material becomes increasingly elastic (or plastically) anisotropic and the converse. At the extreme, if a material is nearly elastically isotropic, assuming the macroscopic stress state is equal to the grain stress state is highly accurate. It was observed for the Ti-7Al, that there was a systematic error in the GNN predictions in both the Source and Target domains. The predictions for the Ti-7Al model are tightly bound around the average stress indicating the model is only learning minor adjustments from an isostress mean-field theory, as opposed to learning full neighborhood effects. However, this is simply a consequence of the fact that any GNN model will converge to an isostress prediction as the material approaches isotropy.

\subsection{Neighborhood Effects}
\label{sec:neighborhood}

\begin{figure}[h]
	\centering
	\includegraphics[width=0.6\textwidth]{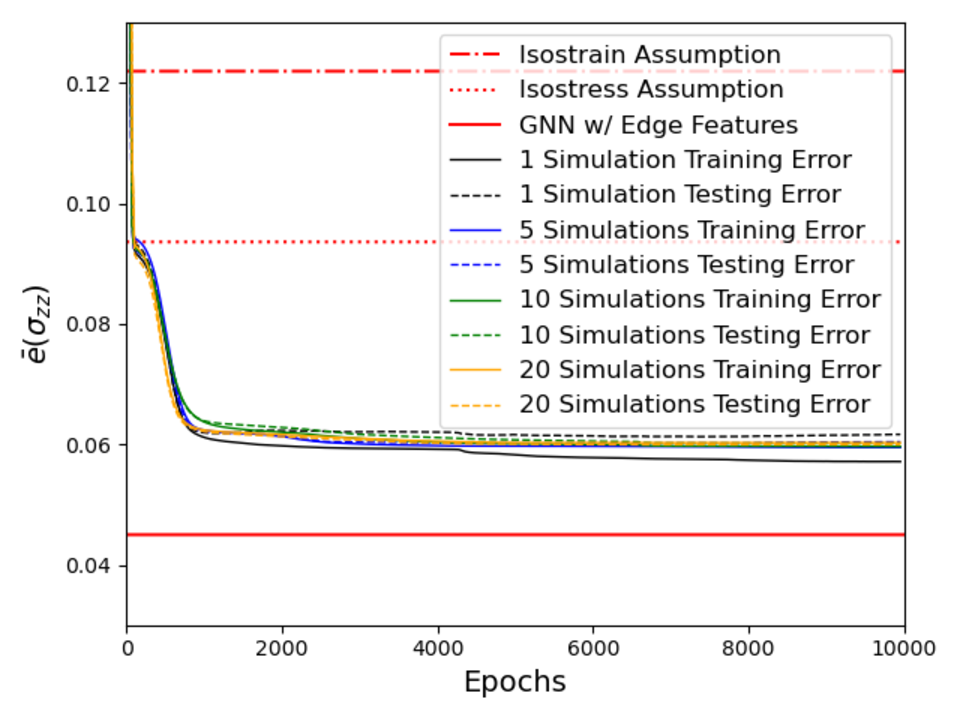}
	  \caption{Learning rates of LSHR grain-scale elastic response using various numbers of CEFEM simulations for training and testing without inclusion of edge features and using the lattice orientation (components of the Rodrigues vector $\bm{r}$) as nodal features.}
	  \label{fig:no_edge}
\end{figure}

It is well established that neighborhood plays a role in the deformation of polycrystals in both the elastic and plastic deformation regime \cite{sarma1996effects,mika1998effects,miller2014understanding} which informed the choice of GNN for polycrystalline modeling and the GM convolution operator selected. Following this, questions arise as to how important are these neighborhood effects in the GNN predictions in this work. As a relatively straightforward means to test the neighborhood effects of the GNN, explicit edge features were removed from the LSHR GNN model formulation and the model was then retrained using the same procedure described above. LSHR was chosen here for its increased anisotropic elasticity. Removing edge features can be physically interpreted as the model knowing the nodal features (orientations) of each neighboring grain, but not necessarily the explicit position of the neighbors with respect to loading. This model formulation lies between that which was used in \S \ref{sec:results} and a more standard mean-field theory which contains no information about neighborhood in its prediction. The results from training and testing against reserved simulations using no edge features and orientation as nodal features (Rodrigues parameterization) are given in Fig.~\ref{fig:no_edge}. Along with isostrain and isostress mean theory predictions included as before, the final accuracy from the full LSHR GNN model described in the results are  included. As would be expected from the physical interpretation of this model without edge features, the accuracy of model prediction lies between that of the full GNN model considering both nodes and edges described above and the isostress prediction. Beyond using five CEFEM simulations for training, the model accuracy stabilizes at approximately 6\% in comparison to approximately 4.5\% for the full GNN model and approximately 10\% for the isostress prediction.

\begin{figure}[h]
	\centering
	\includegraphics[width=1.0\textwidth]{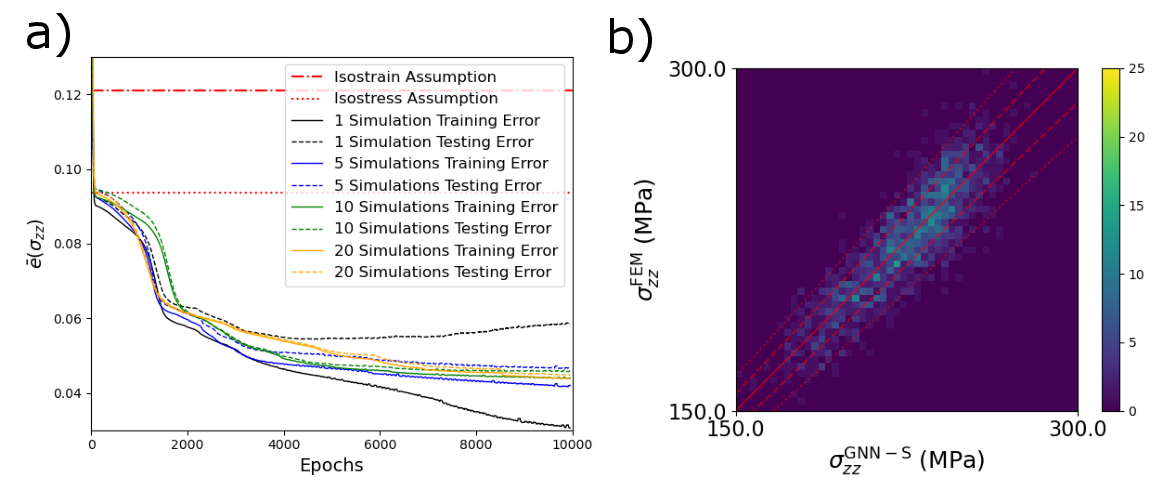}
	  \caption{a) Learning rates of LSHR grain-scale elastic response using various numbers of CEFEM simulations for training and testing generated with a single tessellation and `shuffled' orientation sets using the lattice orientation (components of the Rodrigues vector $\bm{r}$) as nodal features. b) Histogram of the grain scale stress predictions on a new microstructure comparing that from a GNN trained with data from the shuffled orientation sets in a single tessellation (GNN-S) and full-field CEFEM.}
	  \label{fig:shuffle}
\end{figure}

Another question is: how important are the exact grain morphology configurations of a neighborhood for GNN model training? To explore this effect, a single microstructural instantiation of LSHR was generated and the orientations were then shuffled through the fixed grain tessellation. This produces 25 unique microstructures, but with a less diverse number of grain morphologies for training than considered previously. These data were then used to train and test the same GNN model described in \S \ref{sec:results} with lattice orientation as the nodal features (model labeled GNN-S). Learning rates from model training and reserved data testing are given in Fig.~\ref{fig:shuffle}a. The final accuracies at 10,000 epochs for these training data sets are comparable to the more diverse learning set described in the results, approximately 4.5\%. These training and testing results indicate that for the relatively equiaxed microstructures tested in this work, a less diverse pool of grain shape configurations is necessary.  However, to fully evaluate this mode of training with reduced numbers of grain morphology configuration, the GNN-S model was tested on a new tessellation that had also been deformed using CEFEM. Fig.~\ref{fig:shuffle}b shows a histogram comparing the two sets of predictions. The accuracy of the grain-by-grain predictions are comparable to the GNN model trained using a wider array of grain morphology instantiations (see Fig.~\ref{fig:fem_fit}a). We take care to note that with these observations that a limited number of tessellations are necessary for training, and we we do not expect them to continue to be valid for more complex microstructures (e.g., columnar shaped or bimodal grain size distributions) or during plasticity.

\subsection{Extension of Modeling Efforts}

With the ability of the GNNs to predict elastic response shown here, extension of the approach to predicting plasticity should be considered. In this work, prediction of mechanical response of each node is static and does not evolve with time or deformation. To address material evolution that occurs during plastic deformation, recurrent neural networks (RNNs) can be layered on top of the GNN surrogate. In the RNN framework, the neural network has a `memory' that transmits output nodal features through time, enabling dynamic processes to be modeled. This approach has been successful in time-evolving graph problems such as traffic forecasting~\cite{li2018diffusion}, but has mostly been developed for cases where the temporal data is largely stationary (e.g., regular traffic data)
and the nodal features are static (e.g., location information of roads is static). However, existing approaches for modeling plasticity present a path forward as there are significant commonalities between nodal features in a graph and internal state variables characterizing the local microstructure of material points \cite{kocks1975thermodynamics}. One can imagine beyond predicting the mechanical response at nodes after an loading increment (such as this work), nodal features that are either implicit descriptors of microstructure (e.g., slip system strengths) or explicit (e.g., dislocation density) can be predicted.

Once trained, the GNN surrogate models are highly efficient in predicting mechanical response as they do not require the inversion of large systems of equations or any non-linear optimization. This provides an opportunity for embedding into larger-scale models. Efforts to include grain-scale evolution in the larger-scale deformation or forming operations generally requires homogenization \cite{zecevic2017coupling} or a linking hypothesis \cite{beaudoin1995hybrid,dawson2003advances}, usually a Taylor assumption, which will remove local neighborhood effects from consideration, although inclusion of non-explicit neighborhood effects have been attempted \cite{sarma1996texture}. While successful for predicting a bulk property such as anisotropic strength \cite{barton2015use} or a microstructural feature such as phase fraction \cite{zecevic2019crystallographic}, this approach will be unable to predict mechanical responses, such as fracture and fatigue, which are dictated by rare, extreme events. Embedding GNN surrogate models, which naturally incorporate neighborhood effects without a large computational overhead, within a larger-scale simulation is a possible avenue for predicting extreme events in full size components, opening a path for truly microstructurally-sensitive predictions of material failure in component-scale simulations.

\section{Summary}

A novel transfer learning approach was taken for training and evaluating the performance of GNN surrogate models. 
These novel surrogate models provide a new opportunity for implementation in multiscale mechanical modeling where micromechanical response can be rapidly evaluated using a GNN surrogate.
Models were used to predict the elastic response of grains in samples deformed under uniaxial tension. From this work we find that:
\begin{itemize}
    \item GNNs can exceed the accuracy of traditional mean-field theories (models) for predicting anisotropic elastic micromechanical response through incorporation of information regarding local neighborhood.
    \item Micromechanical descriptors such as directional modulus and near equivalent microstructural descriptors such as lattice orientation provide similar accuracy for GNN surrogate model predictions of elasticity.
    \item As mechanical response becomes increasingly isotropic, GNN surrogate model predictions converge toward traditional mean-field predictions.
    \item Using a transfer learning approach, prediction of micromechanical response using GNNs is possible even in scenarios when little training data is available (i.e., experimentally measured microstructures).

\end{itemize}

GNN surrogate modeling, beyond predicting grain-scale elastic response, provides a framework for rapidly predicting more complex processes such as plasticity which can in turn be embedded into larger-scale simulation for prediction of complex material behaviors such as fracture and fatigue.

\section{Acknowledgements}
We would like to thank Dr. Mark Obstalecki and Dr. Paul Shade at the Air Force Research Laboratory for providing the LSHR HEDM data used for GNN model testing. ARB is supported in part by ARO Award MURI, NSF CAREER Award IIS-2045555, and NSF DMS-EPSRC Award 2146079.

\section{Data Availability}
Data will be made available by the authors upon request.

\section{Software Availability}
The GNN model fitting code will be made available upon request. FEPX is available at \url{https://fepx.info/}. Neper is available at \url{https://neper.info/}. HEXRD is available at \url{https://github.com/hexrd}.

\bibliography{bib}

\begin{thebibliography}{10}
\expandafter\ifx\csname url\endcsname\relax
  \def\url#1{\texttt{#1}}\fi
\expandafter\ifx\csname urlprefix\endcsname\relax\def\urlprefix{URL }\fi
\expandafter\ifx\csname href\endcsname\relax
  \def\href#1#2{#2} \def\path#1{#1}\fi

\bibitem{eshelby1957determination}
J.~D. Eshelby, The determination of the elastic field of an ellipsoidal
  inclusion, and related problems, Proceedings of the royal society of London.
  Series A. Mathematical and physical sciences 241~(1226) (1957) 376--396.

\bibitem{marin_a}
E.~Marin, P.~Dawson, On modelling the elasto-viscoplastic response of metals
  using polycrystal plasticity, Computer Methods in Applied Mechanics and
  Engineering 165~(1-4) (1998) 1--21.

\bibitem{lebensohn2012elasto}
R.~A. Lebensohn, A.~K. Kanjarla, P.~Eisenlohr, An elasto-viscoplastic
  formulation based on fast fourier transforms for the prediction of
  micromechanical fields in polycrystalline materials, International Journal of
  Plasticity 32 (2012) 59--69.

\bibitem{hamilton2020graph}
W.~L. Hamilton, Graph representation learning, Synthesis Lectures on Artifical
  Intelligence and Machine Learning 14~(3) (2020) 1--159.

\bibitem{wu2020comprehensive}
Z.~Wu, S.~Pan, F.~Chen, G.~Long, C.~Zhang, S.~Y. Philip, A comprehensive survey
  on graph neural networks, IEEE transactions on neural networks and learning
  systems 32~(1) (2020) 4--24.

\bibitem{zhou2020graph}
J.~Zhou, G.~Cui, S.~Hu, Z.~Zhang, C.~Yang, Z.~Liu, L.~Wang, C.~Li, M.~Sun,
  Graph neural networks: A review of methods and applications, AI Open 1 (2020)
  57--81.

\bibitem{Fuhg2022}
J.~Fuhg, L.~van Wees, M.~Obstalecki, P.~Shade, N.~Bouklas, M.~Kasemer,
  Machine-learning convex and texture-dependent macroscopic yield from crystal
  plasticity simulations, Materialia (2022) 101446.

\bibitem{frankel2020prediction}
A.~Frankel, K.~Tachida, R.~Jones, Prediction of the evolution of the stress
  field of polycrystals undergoing elastic-plastic deformation with a hybrid
  neural network model, Machine Learning: Science and Technology 1~(3) (2020)
  035005.

\bibitem{mianroodi2021teaching}
J.~R. Mianroodi, N.~H~Siboni, D.~Raabe, Teaching solid mechanics to artificial
  intelligence—a fast solver for heterogeneous materials, npj Computational
  Materials 7~(1) (2021) 1--10.

\bibitem{pandey2021machine}
A.~Pandey, R.~Pokharel, Machine learning based surrogate modeling approach for
  mapping crystal deformation in three dimensions, Scripta Materialia 193
  (2021) 1--5.

\bibitem{hamilton2017inductive}
W.~Hamilton, Z.~Ying, J.~Leskovec, Inductive representation learning on large
  graphs, Advances in neural information processing systems 30 (2017).

\bibitem{dai2021graph}
M.~Dai, M.~F. Demirel, Y.~Liang, J.-M. Hu, Graph neural networks for an
  accurate and interpretable prediction of the properties of polycrystalline
  materials, npj Computational Materials 7~(1) (2021) 1--9.

\bibitem{weiss2016survey}
K.~Weiss, T.~M. Khoshgoftaar, D.~Wang, A survey of transfer learning, Journal
  of Big data 3~(1) (2016) 1--40.

\bibitem{neper_paper}
R.~Quey, P.~Dawson, F.~Barbe, Large-scale 3d random polycrystals for the finite
  element method: Generation, meshing and remeshing, Computer Methods in
  Applied Mechanics and Engineering 200~(17-20) (2011) 1729--1745.

\bibitem{renversade}
R.~Quey, L.~Renversade, Optimal polyhedral description of 3d polycrystals:
  Method and application to statistical and synchrotron \protect{X}-ray
  diffraction data, Computer Methods in Applied Mechanics and Engineering 330
  (2018) 308--333.

\bibitem{neper_website}
\href{https://neper.info/}{\protect{Neper: Polycrystal Generation and
  Meshing}}.
\newline\urlprefix\url{https://neper.info/}

\bibitem{geuzaine2009gmsh}
C.~Geuzaine, J.-F. Remacle, Gmsh: A 3-d finite element mesh generator with
  built-in pre-and post-processing facilities, International journal for
  numerical methods in engineering 79~(11) (2009) 1309--1331.

\bibitem{fepx_arxiv}
P.~Dawson, D.~Boyce, \protect{FEpX} -- finite element polycrystals: Theory,
  finite element formulation, numerical implementation and illustrative
  examples (2015).

\bibitem{fepx_website}
\href{https://fepx.info/}{\protect{FEPX: Finite Element Polycrystal
  Plasticity}}.
\newline\urlprefix\url{https://fepx.info/}

\bibitem{quey2022neper}
R.~Quey, M.~Kasemer, The neper/fepx project: Free/open-source polycrystal
  generation, deformation simulation, and post-processing, in: IOP Conference
  Series: Materials Science and Engineering, Vol. 1249, IOP Publishing, 2022,
  p. 012021.

\bibitem{Asaro1985}
R.~Asaro, A.~Needleman, Texture development and strain hardening in rate
  dependent polycrystals, Acta Metallurgica 33~(6) (1985) 923--953.

\bibitem{marin_b}
E.~Marin, P.~Dawson, Elastoplastic finite element analyses of metal
  deformations using polycrystal constitutive models, Computer Methods in
  Applied Mechanics and Engineering 165~(1-4) (1998) 23--41.

\bibitem{roters}
F.~Roters, P.~Eisenlohr, L.~Hantcherli, D.~Tjahjanto, T.~Bieler, D.~Raabe,
  Overview of constitutive laws, kinematics, homogenization and multiscale
  methods in crystal plasticity finite-element modeling: Theory, experiments,
  applications, Acta Materialia 58~(4) (2010) 1152--1211.

\bibitem{turner2012two}
T.~Turner, P.~Shade, M.~Groeber, M.~Miller, M.~Uchic, Two integrated
  experimental and modeling approaches to study strain distributions in nickel
  and nickel-base superalloy polycrystals, Models for Processing and
  Properties, The Minerals, Metals and Materials Society, Seven Springs, PA
  (2012) 643--652.

\bibitem{Fisher1964}
E.~S. Fisher, C.~J. Renken, {Single-crystal elastic moduli and the hcp → bcc
  transformation in Ti, Zr, and Hf}, Physical Review 135~(2A) (1964).

\bibitem{kocks1998texture}
U.~F. Kocks, C.~N. Tom{\'e}, H.-R. Wenk, Texture and anisotropy: preferred
  orientations in polycrystals and their effect on materials properties,
  Cambridge university press, 1998.

\bibitem{poulsen2004three}
H.~F. Poulsen, Three-dimensional X-ray diffraction microscopy: mapping
  polycrystals and their dynamics, Vol. 205, Springer Science \& Business
  Media, 2004.

\bibitem{bernier2011far}
J.~V. Bernier, N.~R. Barton, U.~Lienert, M.~P. Miller, Far-field high-energy
  diffraction microscopy: a tool for intergranular orientation and strain
  analysis, The Journal of Strain Analysis for Engineering Design 46~(7) (2011)
  527--547.

\bibitem{nygren2020algorithm}
K.~E. Nygren, D.~C. Pagan, J.~V. Bernier, M.~P. Miller, An algorithm for
  resolving intragranular orientation fields using coupled far-field and
  near-field high energy x-ray diffraction microscopy, Materials
  Characterization 165 (2020) 110366.

\bibitem{musinski2021statistical}
W.~D. Musinski, P.~A. Shade, D.~C. Pagan, J.~V. Bernier, Statistical aspects of
  grain-level strain evolution and reorientation during the heating and
  elastic-plastic loading of a ni-base superalloy at elevated temperature,
  Materialia 16 (2021) 101063.

\bibitem{pagan2021analysis}
D.~C. Pagan, K.~E. Nygren, M.~P. Miller, Analysis of a three-dimensional slip
  field in a hexagonal ti alloy from in-situ high-energy x-ray diffraction
  microscopy data, Acta Materialia 221 (2021) 117372.

\bibitem{pagan2014connecting}
D.~C. Pagan, M.~P. Miller, Connecting heterogeneous single slip to diffraction
  peak evolution in high-energy monochromatic x-ray experiments, Journal of
  applied crystallography 47~(3) (2014) 887--898.

\bibitem{miller2020understanding}
M.~P. Miller, D.~C. Pagan, A.~J. Beaudoin, K.~E. Nygren, D.~J. Shadle,
  Understanding micromechanical material behavior using synchrotron x-rays and
  in situ loading, Metallurgical and Materials Transactions A 51~(9) (2020)
  4360--4376.

\bibitem{hurley2018characterization}
R.~C. Hurley, E.~B. Herbold, D.~C. Pagan, Characterization of the crystal
  structure, kinematics, stresses and rotations in angular granular quartz
  during compaction, Journal of Applied Crystallography 51~(4) (2018)
  1021--1034.

\bibitem{monti2017geometric}
F.~Monti, D.~Boscaini, J.~Masci, E.~Rodola, J.~Svoboda, M.~M. Bronstein,
  Geometric deep learning on graphs and manifolds using mixture model cnns, in:
  Proceedings of the IEEE conference on computer vision and pattern
  recognition, 2017, pp. 5115--5124.

\bibitem{fey2019fast}
M.~Fey, J.~E. Lenssen, Fast graph representation learning with pytorch
  geometric, arXiv preprint arXiv:1903.02428 (2019).

\bibitem{kipf2016semi}
T.~N. Kipf, M.~Welling, Semi-supervised classification with graph convolutional
  networks, arXiv preprint arXiv:1609.02907 (2016).

\bibitem{sarma1996effects}
G.~B. Sarma, P.~R. Dawson, Effects of interactions among crystals on the
  inhomogeneous deformations of polycrystals, Acta Materialia 44~(5) (1996)
  1937--1953.

\bibitem{mika1998effects}
D.~P. Mika, P.~R. Dawson, Effects of grain interaction on deformation in
  polycrystals, Materials Science and Engineering: A 257~(1) (1998) 62--76.

\bibitem{miller2014understanding}
M.~Miller, P.~Dawson, Understanding local deformation in metallic polycrystals
  using high energy x-rays and finite elements, Current opinion in solid state
  and materials science 18~(5) (2014) 286--299.

\bibitem{li2018diffusion}
Y.~Li, R.~Yu, C.~Shahabi, Y.~Liu, Diffusion convolutional recurrent neural
  network: Data-driven traffic forecasting, in: International Conference on
  Learning Representations, 2018.

\bibitem{kocks1975thermodynamics}
U.~Kocks, A.~Argon, M.~Ashby, Thermodynamics and kinetics of slip, Progress in
  materials science 19 (1975) 1--281.

\bibitem{zecevic2017coupling}
M.~Zecevic, I.~J. Beyerlein, M.~Knezevic, Coupling elasto-plastic
  self-consistent crystal plasticity and implicit finite elements: Applications
  to compression, cyclic tension-compression, and bending to large strains,
  International Journal of Plasticity 93 (2017) 187--211.

\bibitem{beaudoin1995hybrid}
A.~Beaudoin, P.~Dawson, K.~Mathur, U.~Kocks, A hybrid finite element
  formulation for polycrystal plasticity with consideration of macrostructural
  and microstructural linking, International Journal of Plasticity 11~(5)
  (1995) 501--521.

\bibitem{dawson2003advances}
P.~Dawson, S.~MacEwen, P.~Wu, Advances in sheet metal forming analyses: dealing
  with mechanical anisotropy from crystallographic texture, International
  Materials Reviews 48~(2) (2003) 86--122.

\bibitem{sarma1996texture}
G.~B. Sarma, P.~R. Dawson, Texture predictions using a polycrystal plasticity
  model incorporating neighbor interactions, International journal of
  plasticity 12~(8) (1996) 1023--1054.

\bibitem{barton2015use}
N.~R. Barton, J.~V. Bernier, R.~A. Lebensohn, D.~E. Boyce, The use of discrete
  harmonics in direct multi-scale embedding of polycrystal plasticity, Computer
  Methods in Applied Mechanics and Engineering 283 (2015) 224--242.

\bibitem{zecevic2019crystallographic}
M.~Zecevic, M.~V. Upadhyay, E.~Polatidis, T.~Panzner, H.~Van~Swygenhoven,
  M.~Knezevic, A crystallographic extension to the olson-cohen model for
  predicting strain path dependence of martensitic transformation, Acta
  Materialia 166 (2019) 386--401.

\end{thebibliography}
\bibliographystyle{elsarticle-num}

\end{document}